\documentclass[12pt]{article}
\usepackage{jhep-mod-2020}
\usepackage{bm}
\usepackage{amssymb,amsmath,amsthm}
\usepackage{mathrsfs}
\usepackage[utf8]{inputenc}
\usepackage{enumerate}
\hypersetup{colorlinks=true} 
\usepackage{xcolor}
\usepackage{appendix}
\usepackage{graphicx}
\usepackage{dcolumn}
\usepackage{bm}
\usepackage{multirow}
\usepackage{float}
\usepackage{tikz}
\usepackage{colortbl}
\usepackage{cancel}
\usepackage[normalem]{ulem}

\usepackage{tocbasic}
\DeclareTOCStyleEntry[
  beforeskip=.2em plus 1pt,
  pagenumberformat=\textbf
]{tocline}{section}
\definecolor{purple}{rgb}{1,0,1}
\definecolor{lime}{HTML}{A6CE39} 

\newcommand{\orcidicon}{%
	\begin{tikzpicture}
	\draw[lime, fill=lime] (0,0) 
		circle [radius=0.16] 
		node[white] {{\fontfamily{qag}\selectfont \tiny ID}};
	\draw[white, fill=white] (-0.0625,0.095) 
		circle [radius=0.007];
	\end{tikzpicture}	\hspace{-2mm}
}
\newcommand\orcidRaul{{\href{https://orcid.org/0000-0001-6389-6105}{\orcidicon}}}
\newcommand\orcidFrancesco{{\href{https://orcid.org/0000-0002-4727-8953}{\orcidicon}}}
\newcommand\orcidStefano{{\href{https://orcid.org/0000-0002-7632-7443}{\orcidicon}}}
\newcommand\orcidMatt{{\href{https://orcid.org/0000-0003-1088-6485}{\orcidicon}}}
\begin{document}

\title{
{\leftline{\Huge Causal hierarchy in modified gravity}}
}

\author{
\Large Ra\'ul Carballo-Rubio$\,^{1}\orcidRaul$,
Francesco Di Filippo$\,^{2,3,4}\orcidFrancesco$,\\
Stefano Liberati$\,^{2,3,4}\orcidStefano$,
{\sf  and} Matt Visser$\,^{5}$\orcidMatt}

\affiliation{
$^{1}$
Florida Space Institute, University of Central Florida, \\
\null\qquad 12354 Research Parkway, Partnership 1, Orlando, FL, USA
}
\affiliation{
$^{2}$  
SISSA - International School for Advanced Studies, Via Bonomea 265, \\
\null\qquad 34136 Trieste, Italy
}
\affiliation{
$^{3}$ 
IFPU - Institute for Fundamental Physics of the Universe, Via Beirut 2, \\
\null\qquad 34014 Trieste, Italy
}
\affiliation{
$^{4}$ 
INFN Sezione di Trieste, Via Valerio 2, 34127 Trieste, Italy
}
\affiliation{
$^{5}$ 
School of Mathematics and Statistics, Victoria University of Wellington, \\
\null\qquad PO Box 600, Wellington 6140, New Zealand
}

\emailAdd{\\raul.carballorubio@ucf.edu, francesco.difilippo@sissa.it, liberati@sissa.it,
matt.visser@sms.vuw.ac.nz}
\abstract{
\parindent0pt
\parskip7pt

We investigate the causal hierarchy in various modified theories of gravity. In general relativity the standard causal hierarchy, (key elements of which are chronology, causality, strong causality, stable causality, and global hyperbolicity), is well-established. In modified theories of gravity there is typically considerable extra structure, (such as: multiple metrics, aether fields, modified dispersion relations, Ho\v{r}ava-like gravity, parabolic propagation, \emph{etcetera}),  requiring a reassessment and rephrasing of the usual causal hierarchy.  
We shall show that in this extended framework suitable causal hierarchies can indeed be established, and discuss the implications for the interplay between ``superluminal'' propagation and causality. 
The key distinguishing feature is whether the signal velocity is finite or infinite. 
Preserving even minimal notions of causality in the presence of infinite signal velocity requires the aether field to be both unique and hypersurface orthogonal, leading us to introduce the notion of \emph{global parabolicity.}



\medskip
\noindent
D{\sc{ate}}:  Monday 18 May 2020;  \LaTeX-ed \today.

}

\maketitle
\def\d{{\mathrm{d}}}
\section{Introduction}
\label{S:introduction}
In general relativity one can invoke a well-established hierarchy of causality conditions, the key steps in which are chronology, causality, strong causality, stable causality, and global hyperbolicity.
See for instance the pedagogical discussion in references~\cite{hawking-ellis, sachs-wu, wald} and in references~\cite{wiki1,wiki2}. 
See also the more technical discussions in references~\cite{penrose, hawking-sachs, wiki3}, and recent refinements in references~\cite{granada1,granada2}.  The existence of this causal hierarchy underlies and informs the sometimes contentious discussions concerning the interplay between possible superluminal propagation and causality. See for instance references~\cite{Benford:1970, Liberati:2000, Liberati:2001, Barcelo:2004, Bruneton:2006, Cheung:2014}, and specifically references~\cite{Deser:2012, Deser:2013a, Deser:2013b, Deser:2014}. 

Herein we shall seek to generalize the causal hierarchy beyond standard general relativity, to various modified theories of gravity, including multi-metric models,  Einstein-aether models, Ho\v{r}ava-like models, modified dispersion relations, \emph{etcetera}.  In doing so we shall partially revise  and in some cases extend the work of reference~\cite{Sotiriou:2015}, and shall furthermore develop a general notion of curved-spacetime parabolic PDEs, and do so in a manner that still  maintains desirable causality properties. 
Because we wish to keep reasonably close to standard general relativity, we shall focus on ideas that can be closely related to Lorentzian metrics, and shall avoid more general pseudo-Finsler constructions~\cite{Barcelo:2001a,Barcelo:2001b,Visser:2007toy,Skakala:2008a, Skakala:2008b, Skakala:2010,Barcelo:2005}.
Finally, we should point out that modifying the causal hierarchy has a potentially serious impact on at least some classes of ``black hole mimickers''~\cite{Small-dark-heavy, Carballo-Rubio:2018a, Carballo-Rubio:2018b, Carballo-Rubio:2019a, Carballo-Rubio:2019b, Barausse:2020, Zulianello:2020}.

\section{The standard general relativity causal hierarchy}
\label{S:standard}

The standard general relativistic causal hierarchy is pedagogically outlined in many places. For instance, good sources are Hawking--Ellis pages 189--206~\cite{hawking-ellis}, Sachs--Wu pages 257--259~\cite{sachs-wu}, Wald pages 195-199~\cite{wald}, and the discussions in references~\cite{wiki1,wiki2}.
At a more technical level, see Penrose pages 11--38~\cite{penrose} and reference~\cite{hawking-sachs} and the online discussion in reference~\cite{wiki3}.  See also the recent refinements in references~\cite{granada1,granada2}. While overall there is good agreement on final results, sometimes definitions are cryptomorphic --- that is, sometimes definitions and theorems are interchanged. 
There is universal agreement on two of the more basic levels of the causal hierarchy.
\vspace{-10pt}
\begin{itemize}
\itemsep-3pt
\item 
\textbf{Chronology condition}:  There are no closed timelike curves.
\item
\textbf{Causality condition}: There are no closed non-spacelike curves.
\end{itemize}
\vspace{-10pt}
\enlargethispage{30pt}
The absence of closed timelike curves forbids time travel, but already for the closed non-spacelike curves one should sub-divide the discussion into two cases:
\vspace{-10pt}
\begin{enumerate}
\itemsep-3pt
\item[(i)] closed non-spacelike curves with a timelike segment;
\item[(ii)] closed null curves.
\end{enumerate}
\vspace{-10pt}
For closed non-spacelike curves with a timelike segment any observer on the timelike segment would be able to receive a message before it was sent; this is clearly undesirable. Furthermore, any closed  non-spacelike curve with a timelike segment can be deformed into a timelike curve. So closed non-spacelike curves with a timelike segment lead to violation of the chronology condition.

In counterpoint, closed null curves are less \emph{directly} problematic. (Receiving a signal at exactly the same time that it is sent is certainly odd, but not in and of itself logically problematic.) On the other hand, infinitesimal perturbations of closed null curves typically lead to closed non-spacelike curves with a timelike segment (see, \emph{e.g.}, propositions 6.4.4 and 6.4.5 of reference~\cite{hawking-ellis}), which is logically undesirable, and so \emph{indirectly} problematic. For this reason it is generally considered desirable, even at the purely kinematic level, to forbid  closed null curves as well.

\begin{figure}[htbp!]
\begin{center}
\includegraphics[scale=.65]{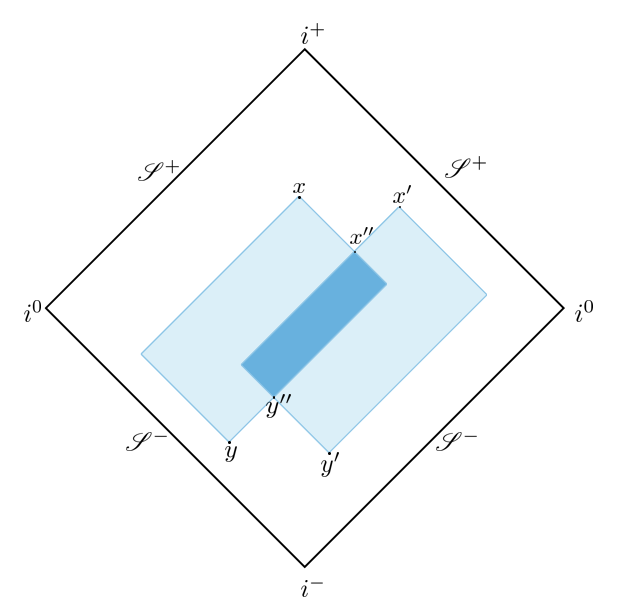}
\caption{The intersection of two chronological diamonds is still a chronological diamond: $I(x,y)\cap I(x',y')=I(x'',y'')$. }
\label{F:diamonds}
\end{center}
\end{figure}

\textbf{Strong causality} is the next step in the usual hierarchy. Perhaps the best way of characterizing stable causality is in terms of the Alexandrov topology based on the chronological diamonds $I(x,y)$.  Assuming that the spacetime is time-orientable, so that one has a meaningful notion of ``future-pointing'', define $I^+(x)$, the chronological future of $x$, as the set of all points that can be reached from $x$ by a future directed timelike curve. 
Similarly, define $I^-(x)$,  the chronological past of $x$, as the set of all points from which one can reach $x$ by following a future directed timelike curve. Then $I(x,y)= I^-(x)\cap I^+(y)$ is the chronological diamond based on $\{x,y\}$. As depicted in figure \ref{F:diamonds}, the intersection of two chronological diamonds is itself a chronological diamond. (See, \emph{e.g.}, Penrose~\cite{penrose} page 33.) 
Thus the set of chronological diamonds can be used as a \emph{basis} for a topology. 
That is, the collection of arbitrary unions of chronological diamonds defines a topology of open sets on spacetime in which a set is open by definition if and only if it is the union of an arbitrary number of chronological diamonds --- this is the Alexandrov topology (the causal topology). 
\vspace{-10pt}
\begin{itemize}
\item \textbf{Strong causality condition}: The Alexandrov topology is Hausdorff.\footnote{A topology is said to be Hausdorff if and only if for any two distinct points $x_1$ and $x_2$, there exist open sets $O_1$ and $O_2$ such that $x_1\in O_1$, $x_2\in O_2$, and $O_1\cap O_2 =\emptyset$.}
\end{itemize}
\vspace{-10pt}
\enlargethispage{10pt}

In fact, the strong causality condition is entirely equivalent to the statement that the Alexandrov topology reproduces the usual manifold topology~\cite{penrose}.   
The strong causality condition can also be rephrased as follows: Surrounding any point $x$ there is an open set $U$ (in the manifold topology) such that any timelike curve that starts at $x$ and then leaves $U$ cannot ever re-enter $U$. (That is: non-spacelike curves cannot return too topologically ``close'' to where they start from.) Overall, the strong causality condition is equivalent to the statement that the light cones allow one to reconstruct the usual spacetime manifold topology. 

Causal diamonds $J(x,y)$, as opposed to chronological diamonds $I(x,y)$, work with non-spacelike curves instead of timelike curves, and lead to closed sets instead of open sets. 
 Define $J^+(x)$, the causal future of $x$, as the set of all points that can be reached from $x$ by a future directed non-spacelike curve. Similarly, define $J^-(x)$,  the causal past of $x$, as the set of all points from which one can reach $x$ by following a future directed non-spacelike curve. Then $J(x,y)= J^-(x)\cap J^+(y)$ is the causal diamond based on the two timelike separated points $\{x,y\}$. 
 
So far we have defined a set of causality conditions which ensure the absence of pathologies in the spacetime we are considering. Given that we are interested in theories where the spacetime is a dynamical object, we can always perform a small perturbation of our spacetime. Therefore we may also want to rule out those spacetimes that are arbitrary close to violating the hierarchy of causality conditions above. To this end, we
construct a partial ordering on the space of Lorentzian metrics $\mathbb{L}(M)$ by saying that one metric $[\hat{g}]_{ab}$ is ``wider'' than another second metric $[g]_{ab}$, denoted $[\hat{g}]_{ab}>[g]_{ab}$, if all non-spacelike vectors in the second metric are strictly timelike in the first metric. This partial ordering can be used (in a completely standard fashion) to define open intervals in the set of all Lorentzian metrics, and these open intervals can be used to define a \emph{sub-basis} for  a topology on the set of Lorentzian metrics one can define on the spacetime manifold --- this topology is typically called the $C^0$ open topology. For the discussion in the next section it will be important to keep in mind that, being a partial order, not every pair of metrics need to be comparable.

With this in mind, the next step in the hierarchy, the \textbf{stable causality condition}, can be defined in at least 3 equivalent ways:
\vspace{-10pt}
\begin{itemize}
\itemsep-3pt
\item 
There exists a global time function $\tau(x)$ whose gradient is everywhere timelike.\\
(So, adopting $-+++$ signature, the vector  $-g^{ab} \nabla_b \tau$ is future-pointing timelike. 
This definition of stable causality implies that for any future-pointing timelike vector $V^a$ one has $V^a \,\nabla_a \tau > 0$.)
\item 
There is a  metric wider than the physical metric such that the wider metric satisfies the {causality} condition.
\item 
There is an open set in the $C^0$ open topology on the set
 of all Lorentzian metrics which contains the physical metric and such that all of the metrics in that open set satisfy the {causality} condition.
\end{itemize}
\vspace{-10pt}
\enlargethispage{20pt}
The second and the third statements are clearly equivalent as they are merely different ways of express the same physical concept. The proof that the first statement is equivalent to the other two  is not trivial, but can be found in standard textbooks, see \emph{e.g.} \cite{wald} for the equivalence between the first and second statement, or \cite{hawking-ellis} for the equivalence between the first and third statement.
(Note that some authors use the strong causality condition instead of the chronology{/causality} condition in the last definition.) 

The last step in the standard causal hierarchy, \textbf{global hyperbolicity}, can also be defined in at least 3 equivalent ways:
\vspace{-10pt}
\begin{itemize}
\itemsep-3pt
\item  
(Causality condition) + (causal diamonds are compact).
\item
Wave equations with suitable initial data have unique solutions.
\item
The spacetime is foliated by spacelike Cauchy hypersurfaces.
\end{itemize}
\vspace{-10pt}

For a technical discussion see references~\cite{wiki3,granada1,granada2}.
We shall soon see that the last of these conditions, the existence of a foliation by suitably redefined and suitably modified Cauchy hypersurfaces, 
is one of the more straightforward causality conditions to work with once one steps far beyond standard general relativity.

Overall the message is this: Precise technical details may differ between various sources, but the basic physics the same. The standard general relativistic causal hierarchy is set up so as to successively exclude various phenomena that for one reason or another might be considered ``unphysical''.  We shall now seek to extend this framework beyond standard general relativity. 

\section{Multi-metric frameworks}
\label{S:multi-metric}

Perhaps the most straightforward extensions of the usual causal hierarchy occur in multi-metric frameworks. 
Multi-metric extensions of general relativity have a long and quite complicated history --- over the last decade one of the key examples of this type of model has been the dRGT ``massive gravity'' models~\cite{deRham:2010a, deRham:2010b, Baccetti:2012a, Baccetti:2012b, Martin-Moruno:2013a, Martin-Moruno:2013b}, though earlier attempts go back several decades~\cite{Visser:1997}. Central to all of these models are multiple Lorentzian metrics $\{[g_i]_{ab}\}_{i=1}^N$ (some dynamical, some possibly non-dynamical background metrics) interacting in various ways.\footnote{Instead of thinking of the $\{[g_i]_{ab}\}_{i=1}^N$ as a set, sometimes it will be useful to think about these metrics as an element $([g_1]_{ab},...,[g_N]_{ab})$ of $\mathbb{L}(M)^N$, the Cartesian product of $N$ copies of $\mathbb{L}(M)$.}

To extend the usual causal hierarchy to such a multi-metric framework at minimum one would want to apply the usual  causal hierarchy to \emph{each} effective metric separately, \emph{and} to demand that you cannot violate causality by switching metrics $[g_i]_{ab}$ part way through whatever physical process you are interested in. Perhaps the simplest way to formulate this is to redefine the notion of chronological/causal curves as follows:
\vspace{-10pt}
\begin{enumerate}
\itemsep-3pt
\item[(i)] 
A piecewise chronological curve is one that can be split into connected segments each one of which is timelike with respect to at least one of the metrics $[g_i]_{ab}$. 
\item[(ii)] 
A piecewise causal curve is one that can be split into connected segments each one of which is non-spacelike with respect to at least one of the metrics $[g_i]_{ab}$. 
\end{enumerate}
\vspace{-10pt}

With these definitions in place we can immediately generalize the chronology and causality conditions in multi-metric framework as follows:
\vspace{-10pt}
\begin{itemize}
\itemsep-3pt
\item 
\textbf{Piecewise chronology condition}:  There are no closed piecewise chronological curves.
\item
\textbf{Piecewise causality condition}: There are no closed piecewise causal curves.
\end{itemize}
\vspace{-10pt}
We can similarly modify the definitions of chronological/causal past/future, and the definitions of chronological/causal diamonds, so as to define strong causality in terms of piecewise timelike curves.

It is useful to break down the piecewise chronology/causality conditions above into equivalent conditions that are often simpler to work with:
\vspace{-10pt}
\begin{itemize}
\itemsep-3pt
\item 
The piecewise chronology condition being satisfied by $\{[g_i]_{ab}\}_{i=1}^N$ is equivalent to each of the individual metrics $[g_i]_{ab}$ independently satisfying the chronology condition, plus the compatibility condition between all the pairs of metrics that $I_i^+(p)\cap I_j^-(p)=\emptyset$, $\forall i,j\in[1,...,N]$.
\item
The piecewise causality condition being satisfied by $\{[g_i]_{ab}\}_{i=1}^N$ is equivalent to each of the individual metrics $[g_i]_{ab}$  independently satisfying the causality condition, plus the compatibility condition between all the pairs of metrics that $J_i^+(p)\cap J_j^-(p)=\emptyset$, $\forall i,j\in[1,...,N]$.
\end{itemize}
\vspace{-10pt}

In the multi-metric framework, 
there are different possible definitions of \textbf{piecewise stable causality} which suitably generalize the usual notion of {stable causality}. 
For instance, working in terms of a global time function:
\vspace{-10pt}
\begin{itemize}
\itemsep-3pt
\item[$A_1:$] 
Each of the individual metrics $[g_i]_{ab}$ is stably causal, and in addition
there exists a global time function $\tau$ whose gradient is everywhere future-pointing with respect to all of the individual metrics $[g_i]_{ab}$.
That is, 
all of the vector fields $[V_i]^a$ that are future-pointing and timelike with respect to the appropriate individual metric $[g_i]_{ab}$ must satisfy $[V_i]^a \,\nabla_a\tau > 0$.
\end{itemize}
\vspace{-10pt}
Thence a piecewise chronological curve, having a tangent vector $V^a$ that is timelike with respect to at least one of the individual metrics $[g_i]_{ab}$, must satisfy $V^a \nabla_a \tau>0$, and so cannot be closed. Thence, this definition implies that all the future-pointing propagation cones must 
lie on the same side of the constant-$\tau$ hypersurfaces.

Furthermore, the two definitions that rely primarily on the existence of the partial order $>$ also generalize naturally, as one just needs to replace $\mathbb{L}(M)$ by $\mathbb{L}(M)^N$, the latter equipped with the product order:
\vspace{-10pt}
\begin{itemize}
\itemsep-3pt
\item[$A_2:$] 
There is an element in $\mathbb{L}(M)^N$ wider than the physical metrics $([g_1]_{ab},...,[g_N]_{ab})$ such that it satisfies the piecewise causal condition.
\item [$A_3:$]
There is an open set in the $C^0$ open topology on the set
 $\mathbb{L}(M)^N$ which contains the physical metrics $([g_1]_{ab},...,[g_N]_{ab})$ and such that all the elements in that open set satisfy the piecewise causal condition.
\end{itemize}
\vspace{-10pt}
One can then show (see Appendix \ref{S:Appendix}) that these definitions are equivalent.

There is an alternative way of proceeding in case the set $\{[g_i]_{ab}\}_{i=1}^N$ satisfies some additional requirements. If there exists a metric in the set $\{[g_i]_{ab}\}_{i=1}^N$ that is wider than the rest, or if there exists a metric $[g_\mathrm{wide}]_{ab}$ in $\mathbb{L}(M)$ that is wider than all the elements $\{[g_i]_{ab}\}_{i=1}^N$ (that is, the propagation cone of $[g_\mathrm{wide}]_{ab}$ contains the union of all the propagation cones of the individual $[g_i]_{ab}$), then we can apply the usual definitions of general relativity the metric $[g_\mathrm{wide}]_{ab}$ :
\vspace{-10pt}
\begin{itemize}
\itemsep-3pt
\item[$B_1:$] 
There exists a global time function whose gradient is everywhere timelike with respect to $[g_\mathrm{wide}]_{ab}$. 
\item[$B_2:$] 
There is a  metric wider than $[g_\mathrm{wide}]_{ab}$ such that the wider metric satisfies the causality condition.
\item[$B_3:$]  
There is an open set in the $C^0$ open topology on the set
 of all Lorentzian metrics which contains the metric $[g_\mathrm{wide}]_{ab}$ and such that all of the metrics in that open set satisfy the causality condition.
\end{itemize}
\vspace{-10pt}
It is also important to notice that, if $[g_\mathrm{wide}]_{ab}$ exists, piecewise chronological/causal curves for $\{[g_i]_{ab}\}_{i=1}^N$ are just ordinary chronological/causal curves of $[g_\mathrm{wide}]_{ab}$.

Hence, we have two triplets of definitions $\{A_i\}_{i=1}^3$ and $\{B_i\}_{i=1}^3$ providing slightly different characterizations of stable causality, which are not equivalent in general. 
It is straightforward to realize that the equivalence between these triplets requires (at a minimum) the existence of $[g_\mathrm{wide}]_{ab}$.
However, there is a situation which, besides from making these two characterizations comparable from the perspective of logical implication, is interesting from a physical standpoint, namely when $\{[g_i]_{ab}\}_{i=1}^N$ is a totally ordered set with respect to $>$. Under this assumption, the two triplets are indeed equivalent, as shown in Appendix~\ref{S:Appendix}. Intuitively, one can see that the necessary condition mentioned above for these being equivalent is satisfied, as $\{[g_i]_{ab}\}_{i=1}^N$ being totally ordered implies that $[g_\mathrm{wide}]_{ab}$ can be identified with the maximal element in $\{[g_i]_{ab}\}_{i=1}^N$.

When it comes to \textbf{global hyperbolicity} all of the 3 standard versions of this notion can easily be adapted to the multi-metric framework, although again there are two possible versions which are not fully equivalent in general:
\vspace{-10pt}
\begin{itemize}
\itemsep-3pt
\item[$C_1:$] 
Causality condition + causal diamonds are compact; (use piecewise causal curves).
\item[$C_2:$] 
For each individual metric  ${[}g_i{]}_{ab}$ wave equations with suitable initial data have unique solutions.
\item[$C_3:$] 
The spacetime is foliated by spacelike Cauchy hypersurfaces; where Cauchy now means spacelike with respect to each individual metric ${[}g_i{]}_{ab}$,  and for each metric  ${[}g_i{]}_{ab}$ the causal curves of that metric intersect the Cauchy surface once and once only.
\end{itemize}
\vspace{-10pt}
Alternatively:
\vspace{-10pt}
\begin{itemize}
\itemsep-3pt
\item[$D_1:$]   
Causality condition + causal diamonds are compact (use $[g_\mathrm{wide}]_{ab}$).
\item[$D_2:$] 
For the metric $[g_\mathrm{wide}]_{ab}$ wave equations with suitable initial data have unique solutions.
\item[$D_3:$] 
The spacetime is foliated by spacelike Cauchy hypersurfaces; where Cauchy now means spacelike with respect to the metric  $[g_\mathrm{wide}]_{ab}$,  and the causal curves of that metric intersect the Cauchy surface once and once only.
\end{itemize}
\vspace{-10pt}
These are the minimum requirements for reformulating the causal hierarchy in a multi-metric framework. It is heartening to see that at least in these multi-metric situations the standard notions of chronology/causality can be successfully adapted without too much violence.

\section{Modified dispersion relations}
\label{S:mdr}

For current purposes we will take ``modified dispersion relations'' to mean the following:
\begin{itemize}
\vspace{-10pt}
\itemsep-3pt
\item For each propagating mode you are interested in, pick a preferred rest frame~\cite{Sotiriou:2011}.
(This does not [for current purposes] necessarily have to be the \emph{same} preferred frame for each propagating mode. More on this point later.)
\item In that preferred rest frame, go to the eikonal approximation to justify writing a dispersion relation $\omega = f(k)$, which will \emph{not be Lorentz invariant} since then it would be an ``ordinary'' dispersion relation~\cite{Sotiriou:2011,Mattingly:2005}. 
\end{itemize}
\vspace{-10pt}
With dispersion relation in hand, the phase and group velocities are defined by
\begin{equation}
v_{phase} = {\omega\over k} = {f(k)\over k}; \qquad\qquad
v_{group} = {\partial \omega\over \partial k} = {f'(k)}.
\end{equation}
A ``signal'' is conventionally defined as an abrupt change in the propagating mode~\cite{waves}, mathematically modelled by a Heaviside step-function, which contains arbitrarily high wavenumbers  in Fourier space. In view of this the ``signal velocity'' can usefully be defined as the infinite wavenumber limit of the phase velocity~\cite{waves}; see also~\cite{Visser:2007toy}.

There are two quite distinct cases:
\vspace{-10pt}
\begin{enumerate}
\itemsep-3pt
\item[(i)] the signal velocity is finite but not the same for all propagating modes;
\item[(ii)] the signal velocity is infinite for at least one of the propagating modes.
\end{enumerate}
\vspace{-10pt}
These two cases lead to rather different causal structure, and significantly different causal hierarchies, as will be discussed in Sections \ref{S:finite} and \ref{S:infinite} below.

Before we move on let us consider several different physical models/frameworks within which the above situations are realized.

\section{Einstein-aether frameworks}
\label{S:aether}
We shall first consider the gravity-aether sector, and then the matter-aether sector. 
\subsection{Gravity-aether sector}
\label{SS:gravity-aether}
\def\L{{\cal L}}
At its most basic,  Einstein-aether theories contain both a metric $g_{ab}$ and a normalized aether field $u^a$ interacting via the most general Lagrangian leading to 2nd-order field equations for both $g_{ab}$ and $u^a$~\cite{Jacobson:2000}:
\begin{equation}
\L = - R - K^{ab}{}_{mn} \;\nabla_a u^m \;\nabla_b u^n - \lambda(g_{ab} u^a u^b +1).
\end{equation}
Here 
\begin{equation}
K^{ab}{}_{mn} = c_1 g^{ab} g_{mn} 
+ c_2 \delta^a{}_m\delta^b_n + c_3 \delta^a{}_n \delta^b{}_m
+c_4 u^a u^b g_{mn}, 
\end{equation}

Since E\"otv\"os type experiments very strongly constrain the universality of free-fall (and so the [weak] equivalence principle) in the most elementary of the  Einstein-aether frameworks one further assumes that the aether field does not directly couple to the matter sector. 
In view of the existence of both metric $g_{ab}$ and aether field $u^a$, one can always define a wider metric $[g_\mathrm{wider}]_{ab} = -E(x)^2 u_a u_b + g_{ab}$, where $E(x)>0$ is (as yet) at one's disposal. 
(The wider metric $[g_\mathrm{wider}]_{ab}$ is often referred to as a \emph{disformal transformation} of the base metric $g_{ab}$. It is also sometimes referred to as a \emph{speed-c metric}, where $c=\sqrt{E^2+1}$.)
One is then temped to immediately impose the standard general relativistic  causal hierarchy on $[g_\mathrm{wider}]_{ab}$, but there is a technical point to investigate first.
How quickly do changes in the aether field propagate through the spacetime?

In reference~\cite{Jacobson:2004}, see also~\cite{Jacobson:2008}, the linearized propagation modes of the combined metric-aether system have been investigated. 
Decomposing into spin-2, spin-1, and spin-0 modes,
the relevant wave equations are all second order, and therefore  for all modes one has $v_{group}=v_{phase}=v_{signal}$. 
Specifically, (see reference~\cite{Jacobson:2008}), we have:
\begin{description}
\item[\qquad Spin-2:] \qquad 
$(v_{\mathrm{signal},2})^2 = {1\over 1 - c_{13}}$.
\item[\qquad Spin-1:] \qquad
$(v_{\mathrm{signal},1})^2 = {2c_1-c_1^2+c_3^2\over 2 c_4(1 - c_{13})}$.
\item[\qquad Spin-0:] \qquad
$(v_{\mathrm{signal},0})^2 = {c_{123}(2-c_{14})\over c_{14}(1 - c_{13})(2+c_{13}+3c_2)}$.
\end{description}
Here as usual $c_{123} = c_1+c_2+c_3$, and similarly for  $c_{13}$ and $c_{14}$. 
The key points are that, barring accidental degeneracies, these signal velocities will be finite, and that these signal velocities are all defined with respect to the \emph{same} aether field/preferred frame.

This makes it straightforward to identify this situation as a particular case of the multi-metric framework discussed in Section \ref{S:multi-metric}. We can then define, in addition to the base metric $g_{ab}$,  three new metrics $[g_i]_{ab} = -E_i^2 u_a u_b + g_{ab}$ where $E_i^2=(v_{\mathrm{signal},i})^2-1$ for $i=0,1,2$. The four element set $\{[g]_{ab}, [g_0]_{ab}, [g_1]_{ab}, [g_2]_{ab}\}$ is totally ordered (which follows from the total order of $\mathbb{R}$ with its standard ordering). Thus, we are precisely in the situation in which the two possible definitions of stable causality and global hyperbolicity in the multi-metric framework are equivalent.

This observation justifies the use of the standard general relativity causal hierarchy, but now applied to the metric $[g_\mathrm{wide}]_{ab}\subset \{[g]_{ab}, [g_0]_{ab}, [g_1]_{ab}, [g_2]_{ab}\}$, in the metric-aether sector of the  Einstein-aether framework.

\subsection{Matter sector}
\label{SS:matter}

Suppose now that in some extended Einstein-aether framework one introduces direct couplings between the matter sector and the aether field.
In view of observational E\"otv\"os type constraints on violations of the equivalence principle,
these direct couplings will certainly be small, but they could in principle be there.
What would this do? 

There is the quite likely possibility that the aether-matter couplings might generically, (either at tree level or due to  higher-order quantum loops), explicitly break Lorentz invariance in the matter sector --- either due to different propagation speeds for different particle species with standard dispersion relation or due to the introduction of higher-spatial-derivative terms,
leading to modified dispersion relations of the type discussed above in section~\ref{S:mdr}. 
One would then get back the dichotomy between finite signal speed and infinite signal speed.
Which of these two options applies depends on specific details of the model, and cannot be decided without a case by case investigation. 
The finiteness \emph{versus} infinitness of the signal speed leads to significantly different causal structures,  see Sections \ref{S:finite} and \ref{S:infinite} below.

\section{Ho\v{r}ava-like frameworks}
\label{S:horava}

The key characteristics of Ho\v{r}ava-like frameworks are the assumed existence of a chronon field (a global time function, often called a khronon field) defining a preferred foliation with respect to which all propagating modes exhibit infinite signal speed.

\subsection{Foliation and signal speed}
\label{S:foliation}

Ho\v{r}ava gravity was developed as an attempt at ameliorating the renormalizability problems of quantum gravity by   violating Lorentz invariance at intermediate stages of the calculation~\cite{Horava:2009a, Horava:2009b, Horava:2010, Horava:2011, Sotiriou:2009a, Sotiriou:2009b,Weinfurtner:2010,Visser:2011}. Effectively one is using Lorentz symmetry breaking as a quantum field theory regulator~\cite{Visser:2009a,Visser:2009b}. 
One keeps 2 time derivatives (to avoid the Ostragowsky instability) but implements at least 6 space derivatives to guarantee power-counting renormalizability. The splitting of spacetime into space+time is taken to be the same for all of the propagating modes, so one is explicitly choosing a preferred foliation of spacetime by spatial 3-surfaces just to set up the formalism. The preferred global time coordinate is typically called the chronon field. 
Thus the key feature of Ho\v{r}ava-like frameworks (in the flat spacetime limit) is the explicit violation of Lorentz invariance leading to PDEs of the form
\begin{equation}
c^{-2} \partial_t^2 \Psi = -c^{-2} m^2  \Psi + \nabla^2\Psi+ H_4 \;\nabla^4\Psi + H_6 \;\nabla^6\Psi +...
\end{equation}

This leads to dispersion relations of the form
\begin{equation}
\omega =  \sqrt{ m^2  + c^2 k^2 - c^2 H_4 \; k^4 + c^2 H_6 \; k^6 + ...}.
\end{equation}
So for the phase velocity we have
\begin{equation}
v_{phase} = {\omega\over k} =\sqrt{ {m^2\over k^2}  + c^2 - c^2 H_4 \; k^2 + c^2 H_6\; k^4+...}
\end{equation}
While for the group velocity
\begin{equation}
v_{group} = {\partial\omega\over\partial  k} =
{ c^2( k - 2H_4 \; k^3 +  3H_6\; k^5+...)
\over 
\sqrt{ {m^2}  + c^2 k^2- c^2 H_4 \; k^4 + c^2 H_6\; k^6+...}}
\end{equation}
Note that
\begin{equation}
{v_{group}\over v_{phase}}  = {k \over \omega} {\partial\omega\over\partial  k} =
{ c^2( k^2 - 2H_4 \; k^4 +  3H_6\; k^6+...)
\over 
{ {m^2}  + c^2 k^2- c^2 H_4 \; k^4 + c^2 H_6\; k^6+...}}
\end{equation}
As long as one is dealing with finite polynomials in the wavenumber, the signal velocity, (the infinite-wavenumber limit of the phase velocity), is now infinite, as is the infinite-wavenumber limit of the group velocity.
So the causal structure is certainly \emph{not} of the usual Lorentzian type.

Note that finite signal speed is typically associated with the Einstein-aether framework rather than the Ho\v{r}ava framework,  especially if one insists on 2nd order field equations both in both the gravity and matter sectors. However, one can also have finite signal speeds with higher-order non-polynomial field equations, for example, when the dispersion relations interpolate between 
 two limit speeds. Notably, an example of this behaviour  can be found in analogue spacetimes based on relativistic BECs~\cite{Fagnocchi:2010}.

Moving away from the flat spacetime limit, each of spatial hypersurfaces in the preferred foliation is a 3-manifold (often called a leaf of the foliation) on which you can construct some Euclidean signature Laplacian $\Delta_3=g^{ac}P_{ab}\nabla^bP_{cd}\nabla^d$, where we have defined the projector $P_{ab}=g_{ab}-V_aV_b$. You can then bootstrap the propagation equations to curved spacetime --- something similar to
\begin{equation}
(V^a \partial_a)^2 \Psi = c^{-2} m^2  \Psi + \Delta_3\Psi+ H_4 (\Delta_3)^2\Psi +...
\end{equation}
Going to the eikonal limit the resulting dispersion relation is polynomial in wavenumber leading to an infinite signal velocity~\cite{Cropp:2013}.
The causal structure is then certainly \emph{not} of the usual Lorentzian type, and the generalized causal hierarchy will be discussed in Section~\ref{S:infinite}. 

\subsection{Relation between Einstein-aether and Ho\v{r}ava frameworks}
\label{S:relation}
The major difference between Einstein-aether and Ho\v{r}ava frameworks is that Einstein-aether theories are based on a preferred threading by integral curves of the aether field $u^a$ whereas Ho\v{r}ava models are based on a preferred foliation by constant chronon hypersurfaces $\tau(x)$. Also, the Einstein-aether framework explicitly asks for second-order equations of motion. Nevertheless there are scenarios where the two formalisms overlap --- one simply needs to drive the vorticity of the aether to zero and to restrict attention to low wavenumbers. 
See particularly reference~\cite{Jacobson:2010}.

\section{Parabolic frameworks}
\label{S:parabolic}

Typical parabolic equations  in flat spacetime are of the form
\begin{equation}
\partial_t \Psi = D \nabla^2 \Psi.
\end{equation}
For instance both the diffusion equation and Schr\"odinger equation are of this form. 
The dispersion relation is of the form $\omega\propto k^2$, leading to an infinite signal velocity. 

How do we generalize this to curved spacetime? 
Assuming the existence of a preferred foliation, with leaves labelled by a chronon field  similar in spirit to that of Ho\v{r}ava-like frameworks, then each of the spatial hypersurfaces is a 3-manifold on which you can construct a Euclidean signature Laplacian $\Delta_3$. 
\begin{itemize}
\itemsep0pt
\item 
By choosing a 4-vector $V^a$ transverse and future pointing with respect to the spatial slices (that is, $V^a \nabla_a \tau  > 0$ you can then bootstrap the parabolic PDEs to curved spacetime:
\begin{equation}
(V^a \nabla_a) \Psi = D  \Delta_3\Psi. 
\end{equation}
Going to the eikonal approximation, the dispersion relation is again of the form $\omega\propto k^2$, again leading to an infinite signal velocity. 
Note that the existence of an infinite signal velocity is intimately related to the preferred foliation.
\item
If in contrast one chooses $V^a \nabla_a \tau  = 0$ then the PDE  reduces to a collection of uncoupled elliptic PDEs, one on each preferred time slice, (one on each leaf of the foliation). 
\item
Finally choosing $V^a \nabla_a \tau  < 0$ results in an \emph{anti-diffusion} equation (a backwards-in-time diffusion equation). 
\end{itemize}

Though the motivation (and many specific details) are now different, the existence of a preferred foliation and infinite signal speed is shared with the Ho\v{r}ava-like models.
The causal structure is certainly \emph{not} of the usual Lorentzian type, and the generalized causal hierarchy will be discussed in Section \ref{S:infinite}. 

Now that we have reviewed the relevant frameworks let us then look  to consider the different scenarios they can lead to for what regards the propagation of signals.

\section{Signal velocities}
In the previous sections we have analyzed three possible frameworks, those arising from Einstein-aether theories, Ho\v{r}ava-like systems, and parabolic equations. The main difference among these frameworks is the speed of propagation of signals being either finite or infinite. An example of the first possibility is Einstein-aether theory. If the aether-matter couplings are such as to still lead to second-order wave equations, then one would still have a set of nested signal cones satisfying $v_{group}=v_{phase}=v_{signal}$ (as above for the metric-aether modes). One would then (as above) define $[g_\mathrm{wide}]_{ab}$ based on the fastest of the signal velocities, and again apply a suitably modified version of the standard general relativity causal hierarchy.

The situation in  Ho\v{r}ava-like and parabolic frameworks where one or more of the signal velocities is infinite is qualitatively different and considerably trickier. (The ``signal metric'' introduced above becomes degenerate, and the ``signal cones'' widen out as far as possible to become ``signal planes'').
Let us now study the two cases of finite and infinite propagation speed in detail.

\subsection{Finite signal velocities}
\label{S:finite}

On general grounds, if all the signal velocities are finite, then the causal hierarchy is extremely similar to that for the multi-metric framework. As per the discussion for modified dispersion relations, pick any propagating mode, and go to the appropriate rest frame. 

Then we can (pointwise) \emph{define} the ``signal metric''
\begin{equation}
ds^2 = - dt^2 + (v_\mathrm{signal})^{-2} \; (dx^2+dy^2+dz^2), 
\end{equation}
with the associated ``signal cones'' $||d\vec x||=v_\mathrm{signal} |dt|$. 
Each propagating mode can be associated with a Lorentzian signal metric of this form, so from a chronology/causality point of view any situation with finite signal velocities can be interpreted as an example of the multi-metric framework. The minor modifications we previously discussed for extending the causal hierarchy to the multi-metric framework will also apply in situations where all the signal velocities are finite. In general situations one can use the definitions $\{A_i\}_{i=1}^3$ and $\{C_i\}_{i=1}^3$ of stable causality and global hyperbolicity, respectively; if $[g_\mathrm{wide}]_{ab}$ exists, one can use instead the definitions $\{B_i\}_{i=1}^3$ and $\{D_i\}_{i=1}^3$ (again, these different prescriptions will be equivalent if the set of metrics is totally ordered).

\subsection{Infinite signal velocities}
\label{S:infinite}

Suppose we have an aether covector field $u_a$ (not at this stage necessarily hypersurface orthogonal) with respect to which some excitation has infinite signal speed. Then:
\begin{itemize}
\vspace{-10pt}
\itemsep-3pt
\item 
Define the analogue of chronological curves in terms of the tangent being future-pointing with respect to the aether, $t^a u_a<0$. (No metric is required to establish this.)

\item
Define the analogue of causal curves in terms of the tangent being non-past-pointing with respect to the aether, $ t^a u_a\leq0$. (No metric is required to establish this.)

\vspace{-10pt}
\end{itemize}
We shall first show that for infinite signal speeds sensible causal behaviour implies that the aether field has to be hypersurface orthogonal. We shall also show that for infinite signal speeds sensible causal behaviour implies that the aether field is unique.

\subsubsection{Hypersurface orthogonality of the aether} 
To demonstrate the need for hypersurface orthogonality of the aether (in the presence of infinite signal velocity)  suppose the the contrary --- that aether is not hypersurface orthogonal,  then its vorticity is nonzero: $\omega^a = \varepsilon^{abcd} u_b u_{[c,d]} \neq 0$. Pick any point $p$ in the manifold and go to Riemann local coordinates, pointing $u^a$ in the  $t$ direction, $(1,0,0,0)$,  and $\omega^a$ in the  $z$ direction, $(0,0,0,1)$. Then
\begin{equation}
g_{ab} = \eta_{ab} + O([\delta x]^2); \qquad
u_a = (-1; \omega y, - \omega x, 0 ) + O([\delta x]^2).
\end{equation}
Here we note $|u|^2 = -1 +  O([\delta x]^2)$ and $\omega^a = (0,0,0,\omega) +  O([\delta x]^2)$.

Now in these coordinates consider the closed topologically circular curve
\begin{equation}
\mathcal{C}: \qquad x^a(\lambda) = 
\Big(0, r_*\cos(\theta(\lambda)), r_* \sin(\theta(\lambda)), 0\Big).
\end{equation}
This curve has tangent vector
\begin{eqnarray}
t^a(\lambda) &=& {dx^a(\lambda)\over d\lambda} 
\nonumber\\
&=& \Big(0, -r_*\sin(\theta(\lambda)), r_* \cos(\theta(\lambda)), 0\Big)\; {d\theta\over d\lambda}
\nonumber\\
&=& r_*\Big(0, -\sin(\theta(\lambda)),  \cos(\theta(\lambda)), 0\Big)\; {d\theta\over d\lambda}. 
\end{eqnarray}
Meanwhile along this curve we have
\begin{equation}
u_a(\lambda) = \Big(-1; \omega r_*  \sin(\theta(\lambda)), - \omega r_*  \cos(\theta(\lambda)), 0 \Big) + O(r_*^2).
\end{equation}

Then
\begin{equation}
t^a u_a = - \omega \;r_*^2 \; {d\theta\over d\lambda}  + O(r_*^3).
\end{equation}
As long as $\omega\neq0$ we can choose $r_*$ small enough to safely ignore the $O(r_*^3)$ term.
Then choose the sign of ${d\theta\over d\lambda}$ to be the same as the sign of $\omega$ and one has a closed chronological curve. Thus if $\omega\neq0$ in the presence of infinite signal velocities the causal hierarchy fails at the very first step.

That is: to preserve the chronology condition in the presence of infinite signal speed the relevant aether field with respect to which infinite signal velocity is defined must have zero vorticity and so be hypersurface orthogonal. So we can set
\begin{equation}\label{eq:time_grad}
u_a = - {\nabla_a \tau\over ||\nabla\tau||}. 
\end{equation}

\subsubsection{Uniqueness of the aether field} 

Suppose now that we have two propagating modes, both with infinite signal speeds, but defined with respect to two distinct hypersurface orthogonal aether fields $u_{1}^a$ and $u_{2}^a$. We want to show that $u_1^a=u_2^a$ if any sensible notion of causality is to survive.
For the current argument it is good enough to work in flat Minkowski space. 
Pick any point $p$ in spacetime and go to coordinates where ${1\over2}(u_1^a+u_2^a)$ is pointing in the $t$ direction, $(1,0,0,0)$,  and the spatial 3-vectors $(u_{1})^i$ and $(u_{2})^i$ are pointing in the $\pm x$ directions, $(\pm1,0,0)$. Then for some $-1<v<1$ we have
\enlargethispage{30pt}
\begin{equation}
g_{ab} = \eta_{ab}; \qquad\qquad
(u_{1,2})_a = \gamma(-1; \pm v, 0,0),
\end{equation}
where $\gamma=(1-v^2)^{1/2}$ so that $|u_{1,2}|^2=-1$.

 Consider the closed circular curve
\begin{equation}
\mathcal{C}: \qquad x^a(\lambda) = 
\Big(0, r_*\cos(\theta(\lambda)), r_* \sin(\theta(\lambda)), 0\Big)
\end{equation}
with tangent vector
\begin{eqnarray}
t^a(\lambda) &=& {dx^a(\lambda)\over d\lambda} \nonumber\\
&=& \Big(0, -r_*\sin(\theta(\lambda)), r_* \cos(\theta(\lambda)), 0\Big)\; 
{d\theta\over d\lambda}
\nonumber\\
&=& r_*\Big(0, -\sin(\theta(\lambda)),  \cos(\theta(\lambda)), 0\Big)\; 
{d\theta\over d\lambda}. 
\end{eqnarray}
Then
\begin{equation}
t^a (u_{1,2})_a =  \pm \gamma\, v\, r_* \; \sin(\theta(\lambda)) \;\; {d\theta\over d\lambda}.
\end{equation}
On each of the two half-circles $\theta\in (0,\pi)$ and $\theta \in(\pi,2\pi)$ the curve will be chronological,  ($t^a u_a <0)$, with respect to either one or the other aether fields. At $\theta=0$ and $\theta=\pi$ the curve is still causal ($t^a u_a \leq0$, in fact $t^a u_a =0$).
But this means one has a closed causal curve with two chronological sections --- so you can send messages into your own past.
The only way to avoid this serious violation of the causality condition is to set $v=0$, that is, for the two aether fields to coincide.

\subsection{Causal structure} 

Hence in the presence of infinite signal velocities the preservation of even the most basic notions of causality implies that there is a global time function $\tau(x)$  whose normalized gradient \eqref{eq:time_grad} is unique and causally well-behaved in the sense described below:
\begin{itemize}
\vspace{-10pt}
\itemsep-3pt
\item 
Define the analogue of chronological curves in terms of the tangent being future-pointing with respect to the global time function, $t^a \nabla_a\tau>0$. \\(No metric is required to do this.)
\item
Define the analogue of causal curves in terms of the tangent being non-past-pointing with respect to the global time function, $t^a \nabla_a\tau\geq0$. \\(No metric is required to do this.)
\vspace{-10pt}
\end{itemize}
With these definitions there automatically are no closed chronological curves --- thence a variant of the chronology condition is built into this formalism. 

There \emph{are} closed causal curves (infinite speed communication) but this is not a problem since by assumption there is a unique global time function to keep things under control.

Using the modified notions of chronology and causality defined above we can still define the notions of chronological past $I^-(x)$ and chronological future $I^+(x)$, but these are no longer cone-like, they are instead half-spaces. One can similarly define ``chronological intervals'' $I(x,y)= I^-(x)\cap I^+(y)$ but these are no longer diamonds, they are instead slabs --- indeed in terms of the preferred global time function one has $I(x,y) = \tau^{-1}\big( (\tau(y),\tau(x) \big)$. There is still an induced Alexandrov topology but is now \emph{not} Hausdorff. (Any 2 distinct points $x$ and $y$ on the same global time slice, $\tau(x)=\tau(y)$, can never be separated by disjoint open sets in this Alexandrov topology since $I^\pm(x)=I^\pm(y)$.) 
\begin{itemize}
\vspace{-10pt}
\itemsep-3pt
\item For infinite signal velocities there is no analogue of strong causality.
\item For infinite signal velocities stable causality needs significant revision. 
\vspace{-10pt}
\end{itemize}
(While by assumption we have a global time function $\tau(x)$ in the most general setting there is not necessarily a Lorentzian metric available to define timelike, and there is certainly by construction no ``wider'' metric to deal with.)
\begin{itemize}
\vspace{-10pt}
\item For infinite signal velocities the concept of global hyperbolicity needs significant revision. 
\vspace{-10pt}
\end{itemize}
(The causality condition is definitely violated and the causal diamonds are no longer diamonds they are slabs; so they are certainly not bounded and certainly not compact.) 

What is instead do-able for infinite signal velocities is to introduce the new concept of \emph{global parabolicity},
which is now defined as demanding a foliation by (parabolic) Cauchy hyper\-surfaces (crossed by chronological curves once and once only). This implies in particular that for any point $x$ we demand that $I^+(x)\cup I^-(x) \cup \tau^{-1}(x)$ is the entire spacetime, so that diffusion equations with suitable initial data (on any fixed but arbitrary slice of the preferred foliation) have unique solutions.

\section{Universal horizons}
\label{S:universal}

Working in Ho\v{r}ava-like models (the key ingredients being the existence of a preferred foliation \emph{and} infinite signal velocity) leads to a new concept --- that of a \emph{universal horizon}~\cite{Eling:2006, Blas:2011,Barausse:2011}; see also references~\cite{Sotiriou:2015,Cropp:2013}.

\begin{figure}[htbp!]
\begin{center}
\includegraphics[scale=.5]{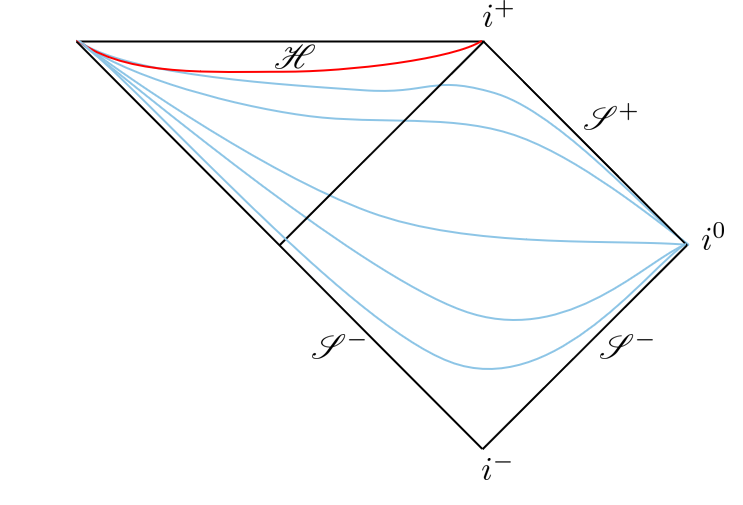}
\caption{Carter--Penrose diagram describing a universal horizon. 
The diagram shows several constant-chronon hyper-surfaces, (leaves of the preferred foliation), with the red line depicting the universal horizon. }
\label{F:carter-penrose}
\end{center}
\end{figure} 

The most basic form of universal horizon arises in static and spherically symmetric situations iff the gradient of the chronon is anti-parallel to the gradient of the radial coordinate, $\nabla_a \tau = - \chi^2 \; \nabla_a r$. They also appear in slowly rotating  black hole solutions in Ho\v{r}ava gravity~\cite{Barausse:2012}. If this happens, then for any chronological curve the tangent vector, by definition satisfying  $t^a \nabla_a\tau>0$, also satisfies $t^a \nabla_a r<0$. Thus all chronological curves are trapped and forced to move ``inwards''. Thence the universal horizon is a constant chronon hypersurface, (a leaf of the preferred foliation), that is simultaneously a constant $r$ hypersurface.  See figure~\ref{F:carter-penrose}. 

To properly define ``static'' one also needs to have a metric $g_{ab}$ available, in order to map the Killing vector into a covector, $K_a = g_{ab}\, K^b$,  to which one can apply the hypersurface orthogonality constraint. One also needs the metric to define the notion of the Killing vector being timelike (sufficiently far away from the black hole region). Under these conditions the static Killing vector is 4-orthogonal to $\nabla r$, and so the universal horizon can equally well be defined by $K^a \nabla_a \tau=0$, see~\cite{Sotiriou:2015}.
Moreover since the static Killing vector is hypersurface orthogonal it induces a natural ``Killing time'' coordinate, $t$, in terms of which we have $K_a \propto \nabla_a t$. 
So at the universal horizon one has $g^{ab} \,\nabla_a \tau \,\nabla_b t = 0$; that is, gradients of the ``chronon time'' and ``Killing time'' are 4-orthogonal (note however that $\nabla_a t$ is spacelike inside the Killing horizon).

Because one has a Lorentzian metric $g_{ab}$ available, one can introduce the usual notions of $i^-$, $i^0$, and $i^+$, and also $\mathscr{J}^\pm$. 
Asymptotically, as one moves ``outwards'' on the constant chronon hypersurface corresponding to the universal horizon one must approach $i^+$. (This is not what one would naively expect in ``normal'' situations, an asymptotic approach to $i^0$. See figure~\ref{F:carter-penrose}. This observation can be modified to develop a general definition of universal horizon.)

\enlargethispage{20pt}
A generic condition for defining a universal horizon is that it is a constant-chronon leaf of the preferred spacetime foliation that contains $i^+$. So a universal horizon would not be a Cauchy hypersurface, since causal curves that intersect $\mathscr{J}^+$ would not necessarily intersect the universal horizon.
In contrast, under ``normal'' conditions constant-chronon leaves of the preferred spacetime foliation asymptote to $i^0$; so they would be Cauchy hypersurfaces.
This compatible with the constructions developed in~\cite{Sotiriou:2015}.

Now is the universal horizon a Cauchy horizon? This depends, very delicately, on precise technical definitions.
The key thing about Cauchy horizons is that something odd is happening to the ``initial data'' needed to define time-evolution into the future.
Is there something odd with universal horizons? 
One issue is this: Since $i^+$ lies on the universal horizon, then anything outside the universal horizon can influence physics at $i^+$.
But then given the assumed infinite signal speeds used to define the universal horizon, \emph{anything} \emph{anywhere} in the domain of outer communication can influence physics anywhere on the universal horizon. This is certainly odd behaviour.

More precisely --- the leaves of the preferred foliation before formation of the universal horizon all asymptote to $i^0$, and so to get a well-defined Cauchy problem at worst one needs to impose some regularity condition at spacelike infinity $i^0$. In contrast, after the universal horizon forms one needs new extra ``initial data'' (corresponding to some regularity condition at future timelike infinity $i^+$) in order to set up a well-defined Cauchy problem. It is in this precise technical sense that the universal horizon can be considered to be a Cauchy horizon. 

There are various ways of rephrasing this in a more formal manner. For instance:  If $x$ lies on a universal horizon then $J^-(x) = J^-(i^+)$. 
Furthermore, if the event $x$ precedes formation of the universal horizon, then $x \in I^-(i^+)$ and $\mathscr{J}^+ \subset I^+(x)$.
The overall message is clear: Preferred foliations combined with infinite signal speeds lead to unusual but internally consistent notions of causal hierarchy. 

\section{The tachyonic anti-telephone}
\label{S:anti-telephome}

Let us now connect the discussion herein back to the famous Benford--Book--Newcomb article on the ``tachyonic anti-telephone''~\cite{Benford:1970}.
What Benford--Book--Newcomb did was to explicitly show that
\begin{equation}
(\hbox{superluminal communication}) + (\hbox{relativity principle}) 
\implies (\hbox{causality violation})
\end{equation}
 Specifically they constructed a closed loop, built out of a combination of  timelike segments and superluminal signals, such that the reply arrived before the query was sent. 
Now turn this logic around:
\begin{equation}
(\hbox{superluminal communication}) + (\hbox{causality}) \implies
(\hbox{extra structure}).
\end{equation}
That is, if on the one hand you want some form of superluminal communication,
and on the other hand you want some sensible notion of causality, then you must have some sort of extra structure that goes beyond standard special relativity or general relativity. 

If (in a ``fundamental'' theory) that extra structure is non-dynamical, then it goes against the grain of Einstein's requirement for ``no prior geometry''.\footnote{There is a messy terminological issue here: 
When Einstein was developing general relativity in the 1910s he was using the (with hindsight) unfortunate phrase ``general covariance''. In modern terminology one distinguishes two separate concepts ``coordinate invariance'' and ``no prior geometry''.
With enough work, it is now realized that \emph{anything} can be made ``coordinate invariant''; one just needs enough non-dynamical background structure.  The physics of what Einstein was trying to get at with his phrase ``general covariance'' was actually what we would now call ``no prior structure'' --- everything (apart from signature, topology, and a few coupling constants) should be dynamical.
}
However in the presence of external constraints (\emph{eg}, the Casimir vacuum between parallel plates) the external constraints provide a natural class of preferred frames.  (This is why, for instance, the Scharnhorst effect~\cite{Liberati:2000,Liberati:2001}, superluminal photon propagation in the Casimir vacuum, is not a problem in terms of causality --- the presence of the parallel plates breaks the 3+1 Lorentz symmetry down to a 2+1 Lorentz symmetry by picking out a preferred direction, the spacelike normal to the plates.)

If that extra structure is dynamical, (be it multi-metric, Einstein-aether, Ho\v{r}ava, or something else), then there are significant observational constraints that should be taken into account.  In short --- the situation is not a free-for-all --- there are tolerably good proposals compatible with good causal structure and effective superluminal signalling, but any such model needs careful phenomenological analysis.

On a more positive note, the presence of such extra structure, (specifically and quite explicitly in the case of imposing a preferred foliation), automatically enforces Hawking's \emph{chronology protection conjecture} thereby keeping the universe safe for historians~\cite{Hawking:1991a,Hawking:1991b,Hawking:2002,Visser:2002,Visser:1992}. 

\enlargethispage{20pt}
\section{Discussion and Conclusions}
\label{S:Conclusions}

The  framework we have developed above allows one to mathematically extend the usual Lorentzian causal hierarchy to multi-metric spacetimes, Einstein-aether models, Ho\v{r}ava-like spacetimes, modified dispersion relations, and parabolic PDEs.  The key dividing point in the analysis is whether the signal velocity is finite or infinite. When the signal speed is finite, a variant of the usual general relativistic causal hierarchy can be formulated.  When the signal speed is infinite, a significantly modified causal hierarchy must be formulated in terms of a global time function (chronon). 
Preserving even minimal notions of causality in the presence of infinite signal velocity requires the aether field to be both unique and hypersurface orthogonal, leading us to introduce the notion of \emph{global parabolicity}.
Either case provides a logically coherent framework for dealing with ``superluminal'' signalling while still maintaining a consistent approach to causality.

\appendix
\section{Proofs of technical propositions in multi-metric frameworks}
\label{S:Appendix}

Let us first show that the three definitions of stable causality in each of the triplets $\{A_i\}_{i=1}^3$ and $\{B_i\}_{i=1}^3$ are equivalent. Dealing with the $\{B_i\}_{i=1}^3$ is straightforward, given that $\{B_i\}_{i=1}^3$ are just the usual definitions for stable causality valid in general relativity but applied to the metric $[g_{\rm wide}]_{ab}$; the equivalence of the three definitions of stable causality within the framework of general relativity is discussed for instance in~\cite{granada1,granada2}. 

Now consider the  $\{A_i\}_{i=1}^3$:
\vspace{-10pt}
\begin{itemize}
\item{$A_1\implies A_2$:} 
The condition $A_1$ implies that all of the metrics $[g_i]_{ab}$ are individually stably causal with respect to individual global time functions $\tau_i$. But we also have the distinct result that for the global time function $\tau$ we can also consider the modified metrics $[\hat g_i]_{ab} = - \lambda_i^2 \,\nabla_a \tau\,\nabla_b\tau + [g_i]_{ab}$ with $\lambda_i\neq0$. These metrics can more formally be written as $[\hat g_i] = -\lambda_i^2 (d\tau)^2 + [g_i]$. For any vector $u^a$ these metrics  will satisfy $[\hat g_i]_{ab} u^a u^b = - \lambda_i^2\, (u^a\nabla_a \tau)^2 + [g_i]_{ab} u^a u^b <   [g_i]_{ab}\, u^a u^b$, and so these modified metrics are all wider than than the original metrics $[g_i]_{ab}$.
Furthermore, by assumption we know that for vectors $[V_i]^a$ that are future-pointing timelike with respect to $[g_i]_{ab}$ we have $[V_i]^a\,\nabla_a\tau > 0$. But then, by continuity,  for vectors  $[\hat V_i]^a$ that are future-pointing timelike with respect to $[\hat g_i]_{ab}$ we can choose $\lambda_i$ to be sufficiently small that we also have $[\hat V_i]^a\,\nabla_a\tau > 0$, while maintaining the condition $[\hat{g}_i]_{ab} > [g_i]_{ab}$.
But this now forbids the existence of closed piecewise causal curves. 
Hence, there exists an element $([\hat{g}_1]_{ab},...,[\hat{g}_N]_{ab})\in \mathbb{L(}M)^N$ wider than $([g_1]_{ab},...,[g_N]_{ab})$ satisfying the piecewise causal condition.

\item{$A_2\implies A_1$:
First note that the $A_2$ version of piecewise stable causality implies in particular the existence of metrics $[\hat g_i]_{ab} > [g_i]_{ab}$ that individually satisfy the usual causality condition. 
Thus usual stable causality holds for each of the individual metrics $ [g_i]_{ab}$, which we can exploit in order to guarantee that each individual $[g_i]_{ab}$ admits a global time function $\tau_i$~\cite{granada1,granada2}. Additionally, the individual metrics in $([\hat{g}_1]_{ab},...,[\hat{g}_N]_{ab})\in L(M)^N$ satisfying $([\hat{g}_1]_{ab},...,[\hat{g}_N]_{ab})>([g_1]_{ab},...,[g_N]_{ab})$ must verify the compatibility condition $\hat{J}_i^+(p)\cap \hat{J}_j^-(p)=\emptyset$, $\forall i,j\in[1,...,N]$. 
}

Geometrically the condition $\hat{J}_i^+(p)\cap \hat{J}_j^-(p)=\emptyset$ means that none of the past propagation cones of any of the $[g_i]_{ab}$ can intersect any of the future propagation cones of any of the $[g_i]_{ab}$.  This implies that all of the future propagation cones must lie on the same side of some hypersurface with normal $n_a\propto \nabla_a\tau$. (We do not need to explicitly find this hyperplane, we just need to know that it exists.) Then for all vectors $[V_i]^a$ that are future-pointing timelike with respect to the respective metric $[g_i]_{ab}$ we have $[V_i]^a\; n_a > 0$ and so $[V_i]^a \;\nabla_a \tau > 0$.

\item{$A_2\;\Longleftrightarrow\; A_3$: This implication and its converse are as straightforward as they are in standard general relativity~\cite{granada1,granada2}. The existence of elements of  $\mathbb{L}(M)^N$ such that  $([\hat{g}_1]_{ab},...,[\hat{g}_N]_{ab})>([g_1]_{ab},...,[g_N]_{ab})$, while both satisfying the piecewise causality condition implies the existence of an open set in $\mathbb{L}(M)^N$ satisfying the piecewise causality condition.
Conversely, since the open sets in $\mathbb{L}(M)^N$ are defined using the partial order $>$ defined on $\mathbb{L}(M)^N$, the open sets are just unions of intervals defined by this partial order. Consequently if $([g_1]_{ab},...,[g_N]_{ab})$ lies in an open set, it must lie within at least one such interval defined by this partial order. The maximal element in this interval will then satisfy both the piecewise causality condition and the inequality $([\hat{g}_1]_{ab},...,[\hat{g}_N]_{ab})>([g_1]_{ab},...,[g_N]_{ab})$.
}
\end{itemize}

\enlargethispage{40pt}
Let us now show that the two triplets $\{A_i\}_{i=1}^3$ and $\{B_i\}_{i=1}^3$ are equivalent when $\{[g_i]_{ab}\}_{i=1}^N$ is a totally ordered set. More specifically, we can show that:
\vspace{-10pt}
\begin{itemize}
\item{$A_2\implies B_2$: If there exists $([\hat{g}_1]_{ab},...,[\hat{g}_N]_{ab})>([g_1]_{ab},...,[g_N]_{ab})$ and the set $\{[g_i]_{ab}\}_{i=1}^N$ is totally ordered, then we can identify $[g_{\rm wide}]_{ab}=[g_j]_{ab}$ as the maximal element in $\{[g_i]_{ab}\}_{i=1}^N$. On the other hand, the piecewise causality condition satisfied by $\{[\hat{g}_i]_{ab}\}_{i=1}^N$ implies the causality condition for each of these metrics considered individually.  In particular, there is at least one $[\hat{g}_j]_{ab}>[g_{\rm wide}]_{ab}$ which will satisfy the causality condition.}
\item{$B_2\implies A_2$: Again, we can identify $[g_{\rm wide}]_{ab}=[g_j]_{ab}$ as the maximal element in $\{[g_i]_{ab}\}_{i=1}^N$. We exploit the assumed existence of a metric $[\hat{g}_{\rm wide}]_{ab}>[g_{\rm wide}]_{ab}$ satisfying the causality condition in order to construct $([\hat{g}_{\rm wide}]_{ab},...,[\hat{g}_{\rm wide}]_{ab})\in \mathbb{L}(M)^N$, which trivially satisfies the piecewise causality condition, as well as satisfying $([\hat{g}_{\rm wide}]_{ab},...,[\hat{g}_{\rm wide}]_{ab})>([g_1]_{ab},...,[g_N]_{ab})$.}
\end{itemize}
Finally we note that in general relativity, stable chronology (stability of the chronology condition under $C^0$ perturbations) and stable causality are often used interchangeably due to their equivalence~\cite{granada1,granada2}. For completeness, we can show that a similar statement holds for their piecewise generalizations in a multi-metric framework:
\begin{itemize}
\item{\emph{Equivalence of piecewise stable chronology and piecewise stable causality}: It is clear that the latter implies the former. To tackle the reverse implication, let us show that if a spacetime is not piecewise stably causal, then it cannot be piecewise stably chronological. In the $C^0$ open topology, pick an open set surrounding $\{[{g}_i]_{ab}\}_{i=1}^N$. Within that open set, pick metrics  $[\hat{g}_i]_{ab} > [\tilde{g}_i]_{ab} > [g_i]_{ab}$ for each $i$ and for any choice of $\{[\hat{g}_i]_{ab}\}_{i=1}^N$. Then the spacetime not being piecewise stably causal implies that you can choose the $\{[\tilde{g}_i]_{ab}\}_{i=1}^N$ so as to permit the existence of at least one closed causal curve for the $\{[\tilde{g}_i]_{ab}\}_{i=1}^N$. That closed curve must be piecewise timelike for $\{[\hat{g}_i]_{ab}\}_{i=1}^N$, so that the latter cannot satisfy the piecewise chronological condition.
}
\end{itemize}
%

\subsection*{Acknowledgments}
RCR acknowledges support from the Preeminent Postdoctoral Program (P3) at UCF.\\
SL and FDF  acknowledge funding from the Italian Ministry of Education and  Scientific
Research (MIUR)  under the grant  PRIN MIUR 2017-MB8AEZ.\\
MV was supported by the Marsden Fund, via a grant administered by the Royal Society of New Zealand. 


\end{document}